\documentstyle[prl,aps,epsf,multicol]{revtex}

\begin{document}

\draft

\newcommand{\npa}[1]{Nucl.~Phys.~{A#1}}
\newcommand{\etal}{{\em et al.}}

\title{Differential Elliptic Flow in
2~-~6 AGeV Au~+~Au Collisions:\\
A New Constraint for the Nuclear Equation of State}

\author{
P.~Chung$^{(1)}$, N.~N.~Ajitanand$^{(1)}$, J.~M.~Alexander$^{(1)}$, J.~Ames$^{\dagger}$
M.~Anderson$^{(5)}$, D.~Best$^{(2)}$,
F.P.~Brady$^{(5)}$, T.~Case$^{(2)}$, W.~Caskey$^{(5)}$, D.~Cebra$^{(5)}$,
J.L.~Chance$^{(5)}$, B.~Cole$^{(11)}$, K.~Crowe$^{(2)}$,
A.~C.~Das$^{(3)}$, J.E.~Draper$^{(5)}$, M.L.~Gilkes$^{(1)}$,
S.~Gushue$^{(1,9)}$, M.~Heffner$^{(5)}$,
A.S.~Hirsch$^{(7)}$, E.L.~Hjort$^{(7)}$, W.~Holzmann$^{(1)}$, L.~Huo$^{(13)}$, M.~Issah$^{(1)}$,
M.~Justice$^{(4)}$,
M.~Kaplan$^{(8)}$, D.~Keane$^{(4)}$, J.C.~Kintner$^{(12)}$, J.~Klay$^{(5)}$,
D.~Krofcheck$^{(10)}$, R.~A. ~Lacey$^{(1)}$, J.~Lauret$^{(1)}$, M.A.~Lisa$^{(3)}$, H.~Liu$^{(4)}$, Y.M.~Liu$^{(13)}$,
J.~Milan$^{(1)}$, R.~McGrath$^{(1)}$, Z.~Milosevich$^{(8)}$,
G.~Odyniec$^{(2)}$, D.L.~Olson$^{(2)}$,
S.~Panitkin$^{(4)}$, C.~Pinkenburg$^{(9)}$, N.T.~Porile$^{(7)}$, G.~Rai$^{(2)}$, H.G.~Ritter$^{(2)}$,
 J.L.~Romero$^{(5)}$, R.~Scharenberg$^{(7)}$, L.~Schroeder$^{(2)}$,
B.~Srivastava$^{(7)}$, N.T.B~Stone$^{(2)}$, T.J.M.~Symons$^{(2)}$,
J.~Whitfield$^{(8)}$, T.~Wienold$^{(2)}$, R.~Witt$^{(4)}$, L.~Wood$^{(5)}$,
and W.N.~Zhang$^{(13)}$
                        \\  (E895 Collaboration ) \\
P.~Danielewicz$^{(6)}$ }

\address{
$^{(1)}$Depts. of Chemistry and Physics,
SUNY at Stony Brook, New York 11794-3400 \\
$^{(2)}$Lawrence Berkeley National Laboratory,
Berkeley, California, 94720\\
$^{(3)}$Ohio State University, Columbus, Ohio 43210\\
$^{(4)}$Kent State University, Kent, Ohio 44242 \\
$^{(5)}$University of California, Davis, California, 95616 \\
$^{6}$ Michigan State University, East Lansing MI 48824-1321 and \\
Gesellschaft f\"ur Schwerionenforschung, Darmstadt, 64291 Germany\\
$^{(7)}$Purdue University, West Lafayette, Indiana, 47907-1396 \\
$^{(8)}$Carnegie Mellon University, Pittsburgh, Pennsylvania 15213\\
$^{(9)}$Brookhaven National Laboratory, Upton, New York 11973 \\
$^{(10)}$University of Auckland, Auckland, New Zealand \\
$^{(11)}$Columbia University, New York, New York 10027 \\
$^{(12)}$St. Mary's College, Moraga, California  94575 \\
$^{(13)}$Harbin Institute of Technology, Harbin, 150001 P.~R. China \\
 }
\date{\today}
\maketitle
%
%
\begin{abstract}

Proton elliptic flow is studied as a function of impact-parameter $b$,
for two transverse momentum cuts in 2~-~6 AGeV Au~+~Au collisions. 
The elliptic flow shows an essentially linear dependence 
on b (for $1.5 < b < 8$ fm) with a negative slope at 2 AGeV,
a positive slope at 6 AGeV and a near zero slope at 4 AGeV.
These dependencies serve as an important constraint for discriminating between 
various equations of state (EOS) for high density nuclear matter, and they
provide important insights on the interplay between collision 
geometry and the expansion dynamics. Extensive comparisons of the measured 
and calculated differential flows provide further evidence for a softening of the 
EOS between 2 and 6 GeV/nucleon.
\end{abstract}

\begin{multicols}{2}
\narrowtext

        For several years now, the study of nuclear matter at high energy
density has held the promise of providing valuable insights on the
nuclear equation of state (EOS) and on the predicted phase
transition to a quark-gluon plasma~(QGP)\cite{stocker86,qm96,reisdorf97}.
At~AGS energies of $\sim 1 - 14$ AGeV, elliptic flow has emerged
as an invaluable probe of high density nuclear matter\cite{sor97,bar97,dan98,pin99}.
This flow has been attributed to a delicate balance between (i)
the ability of compressional pressure to effect a rapid transverse
expansion of nuclear matter and (ii) the passage time
for removal of the shadowing of participant hadrons by the
projectile and target spectators\cite{dan95}.
If the passage time is long compared to the expansion time, spectator
nucleons serve to block the path of participant hadrons emitted toward
the reaction plane, and nuclear matter is squeezed-out
perpendicular to this plane giving rise to negative elliptic flow.  For shorter 
passage times, the blocking of participant matter is significantly reduced and 
preferential in-plane emission or positive elliptic flow is favored because the geometry
of the participant region exposes a~larger surface area in the direction
of the reaction plane. Thus, elliptic flow is predicted and found to be
negative for beam energies $\lesssim 4$~AGeV and positive for higher
beam energies\cite{dan98,pin99,oll92}.

        Recent theoretical studies of elliptic flow have suggested
a sensitivity to the pressure at maximum compression~\cite{sor97,oll92,oll93} and thus to
the stiffness of the EOS, and to possible QGP formation\cite{dan98}.
Despite this sensitivity, the commonly calculated patterns for
elliptic flow very often do not constrain the EOS uniquely.
This being the case, it is important to investigate additional
experimental observables which may provide more stringent constraints for the EOS.
Here, we investigate the utility of differential elliptic flow 
measurements $v_2(b)$ and $v_2(b,p_T)$, as possible constraints.

    The measurements were performed at  the Alternating Gradient
Synchrotron~(AGS) at the Brookhaven National Laboratory.  Beams of $^{197}$Au
at $E_{Beam} = 2$, 4, and 6~AGeV\cite{beam_energy} were used to bombard a
$^{197}$Au target of thickness calculated for a 3\% interaction probability.
Typical beam intensities  resulted in $ \sim 10$ spills/min with $\sim 10^3$  
particles per spill.  Charged reaction products were detected in the Time 
Projection Chamber (TPC)\cite{GRai90} of the E895 experimental setup.  The TPC
located in the MPS magnet (typically at 1.0 Tesla)  provided good
acceptance and charge resolution for charged particles $-1<Z<6$ at all three
beam energies. However, a unique mass resolution for $Z=1$
particles was not achieved for all rigidities. Data were taken in two experimental
runs with a trigger which allowed for a wide range of impact-parameter
selections as presented below.

         Our flow analysis follows the now standard
procedure~\cite{dem90} of using the
second Fourier coefficient, $v_2 = \langle \cos{2 \phi}
\rangle $, to measure the elliptic flow or asymmetry of the proton
azimuthal distributions at mid rapidity ($|y_{cm}| < 0.1$).  This
distribution can be expanded as
\begin{equation}
 {dN \over d\phi} \propto \left[ 1 + 2\,v_1\cos(\phi)
+2\, v_2\cos(2\phi)  \right] \, , \label{Dist}
\end{equation}
where $\phi$ represents the azimuthal angle of an emitted proton relative to
the reaction plane.  Near mid rapidity in a symmetric system
$v_1 \approx 0$.  The reference
azimuthal angle $ \Phi_{plane}$ of the reaction plane is
determined using\cite{dan85} the vector
{\bf Q$_i$} = $\sum_{j\neq i}^{n}{w(y_j) \, {{\bf p_T}_j  /
{p_T}_j }}$.
Here,  ${\bf p_T}_j$ and $y_j$ represent,  respectively,
the transverse momentum and the rapidity of baryon~j (Z$\le 2$) in an event.
The weight $ w (y_j)$ is assigned the  value ${ <p_x>/<p_T>}$, where
$p_x$ is the transverse momentum in the reaction
plane\cite{dan95}. The average $<p_x>$ is
obtained from an earlier pass of an iterative procedure
employed for each energy and impact-parameter selection.

The orientation of the impact-parameter vector follows
azimuthal symmetry about the beam axis.
Therefore, the azimuthal distribution of the determined reaction plane
should be uniform or flat. We have established that
deviations from this uniformity can be attributed to deficiencies
in the acceptance of the TPC and have applied rapidity and
multiplicity dependent corrections following Ref.~\cite{pin99}.
The corrections were applied for each of several
impact-parameter
selections at each beam energy; they ensure the absence of spurious
elliptic flow signals which might result from distortions in
the reaction plane distribution.
The dispersion of the reaction plane $<|\phi_{12}|>/2$ was estimated
for each impact-parameter $b$ via the sub-event
method\cite{dan85}.
Suffice to say, a reasonable resolution was observed over the entire range of energy
and $b$ studied. These estimates for the reaction plane
dispersion serve as a basis for evaluating the dispersion corrections summarized in Table~1;
these corrections have been applied to the extracted flow values discussed below.

The event multiplicity of identified charged
particles $M_{filt}$ was used for centrality selection.
That is, several multiplicity bins were selected in the range
from 0.4 to 1.0 $M_{max}$ where $M_{max}$ is the
point in the charged particle multiplicity distribution 
where the height of the distribution has fallen to
half its plateau value\cite{gut89}. Impact-parameter estimates have also
been made for these centrality selections, at each beam energy,
via their respective fraction of the minimum bias cross section.

\begin{figure}
\centerline{\epsfysize=3.8in \epsffile{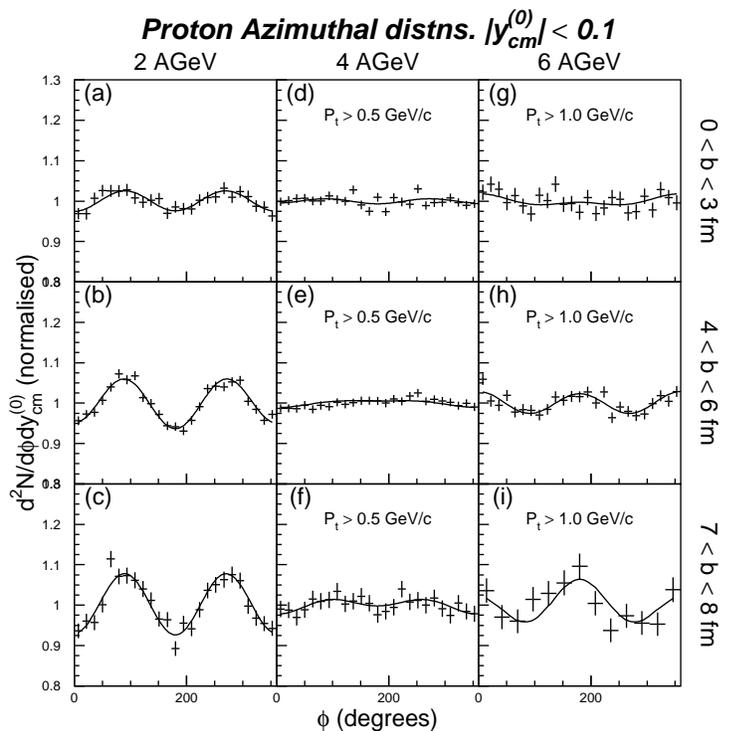}}
\caption{ Measured azimuthal distributions for Au~+~Au collisions.
Distributions are shown for the impact-parameter ranges of
$0\leq b \leq 3$ fm, $4\leq b \leq 6$ fm
and $7\leq b \leq 8 fm$ and the beam energies of 2 (a, b,
c), 4 (d, e, f) and 6 (g, h, i) AGeV, as indicated. The solid lines are
drawn to guide the eye.
}
\label{azi_dist}
\end{figure}

        Figure~\ref{azi_dist} shows
representative distributions in the azimuthal angle $\phi $
obtained at the energies of 2, 4
and 6 AGeV for mid-rapidity ($|y^{(o)}_{cm}|< 0.1$) protons.
The panels from left to right represent the three beam
energies, respectively, and from top to bottom
the three impact-parameter ranges of
$0 \lesssim b \lesssim 3$, $4 \lesssim b \lesssim 6$
and $7 \lesssim b \lesssim 8$  fm.  For visual clarity,
a $p_T$ cut has been applied to the distributions shown
for both the 4 and 6 AGeV data, as indicated.
Within each $b$-range in Fig.~\ref{azi_dist}, the previously reported
transition from negative to positive elliptic flow
at $\approx 4$ AGeV\cite{pin99} is clearly seen.
That is, the elliptic flow is negative at 2 AGeV, positive at 6 AGeV and
essentially zero at 4 AGeV.  An apparent increase of the
anisotropy of the distributions with increasing $b$ can also be discerned for
the 2 and 6 AGeV data shown in Fig.~\ref{azi_dist}. We attribute this
trend to an interplay of the changing geometry with the
expansion of excited participant matter as discussed below.

\begin{figure}
\centerline{\epsfysize=5.2in \epsffile{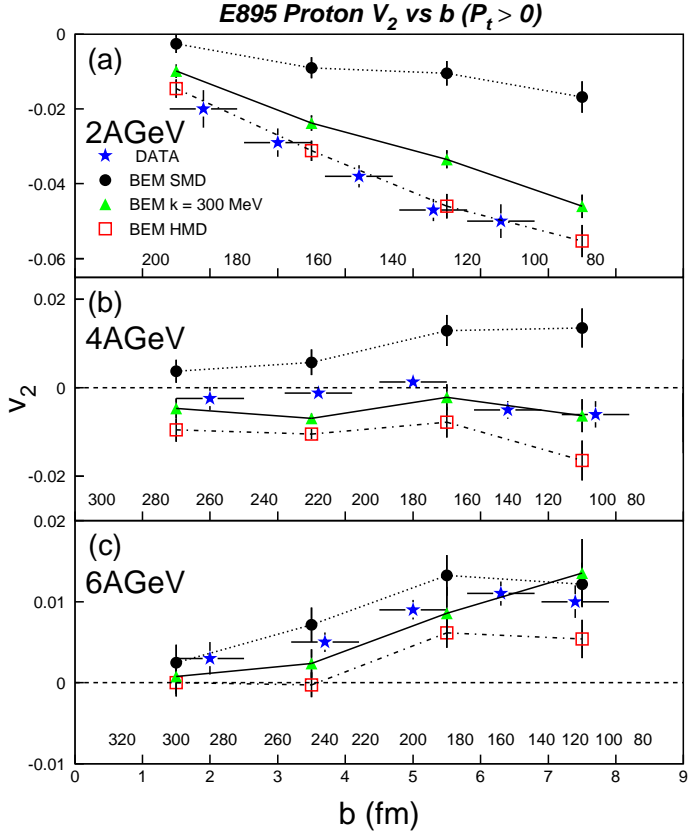}}
\vspace*{-0.2in}
\caption{$v_2$ as a function of $b$ ($p_T > 0$) for 2
(a), 4 (b) and 6 (c) AGeV Au~+~Au collisions.
Experimental values are indicated by the filled stars. 
The open squares, full circles and solid triangles represent
$v_2$ values from BEM calculations with a stiff
($K~=~380$ MeV), a soft ($K~=~210$ MeV) and
an intermediate ($K~=~300$ MeV) momentum-dependent EOS respectively.
The identified charged particle multiplicity $M_{filt}$, is also indicated for 
several values of $b$. The solid, dotted and dashed-dotted lines serve to guide the
eye only.
}

\label{bdep-v2}
\end{figure}

        Figure~\ref{bdep-v2} shows the $v_2$ coefficients for the 
full $p_T$ range, as a function of $b$ for data (filled stars) obtained 
at 2, 4, and 6 AGeV in the three panels, respectively.
These coefficients have been obtained by evaluating 
the $\langle \cos{2 \phi}\rangle$ for each azimuthal
distribution obtained for a given impact-parameter at
each beam energy. A correction has been applied to some of
these coefficients to account for biases resulting from 
(i) low $p_T$ acceptance losses in the TPC for the 2, 4, and 6 AGeV beams,
(ii) high $p_T$ acceptance losses in the TPC for the 2 AGeV beam, and 
(iii) $\pi^+$ contamination of the proton sample at 4 and 6 AGeV\cite{pid_mixup,pin99}.
A procedure for effecting these corrections has been detailed 
in Ref.\cite{pin99}. That is, we first plotted the
observed Fourier coefficient $\langle \cos{2 \phi } '\rangle $~vs.~$p_T$ with
$p_T$ thresholds which allowed clean particle separation ($p_T \sim 1$~GeV/c).
We then extracted the coefficients for the quadratic dependence of  $\langle
\cos{2 \phi } '\rangle $ on ~$p_T$. These quadratic fits
 are restricted by the requirement  that $\langle \cos{2 \phi } '\rangle = 0$
for  $p_T = 0$.  Next, we corrected the proton $p_T$ distributions for possible
high and low $p_T$ losses. 
A weighted average (relative number of protons in a $p_T$
bin times the $\langle \cos{2 \phi } '\rangle $ for that bin) was then
performed to obtain $\langle \cos{2 \phi } '\rangle $ for each beam energy.
The corrections which result from this procedure are $\sim$ 5\% for the 4 and 6 AGeV 
beams and $\sim$ 15\% for the 2 AGeV beam.
Subsequent to these evaluations, the $v_2$ values were corrected
for reaction plane dispersion using the procedures detailed in
Refs.~\cite{pin99,dem90,dan85,oll97}.

        The $v_2$ values represented by filled stars in
Fig.~\ref{bdep-v2} indicate an essentially linear dependence 
on impact parameter.  The slope of this dependence is 
clearly negative and positive for the 2 and 6 AGeV
data, respectively. By contrast, an essentially flat dependence
is observed for the 4 AGeV data suggesting that the beam energy at which the
elliptic flow changes sign is not very sensitive to $b$ 
for $0 \lesssim b \lesssim 8$fm.
The approximately linear dependence exhibited by the data can be understood
in terms of the collision geometry and the development of
transverse expansion within the participant matter.  At 2 AGeV
an expansion perpendicular to the reaction plane develops over
the characteristic time of $d/c_s$ while the spectators are
present. Here, $c_s =\sqrt{\partial p / \partial e}$
represents the speed of sound for a given pressure $p$,
and energy density $e$, and $d$ is the perpendicular distance from the
center of the participant region to the surface. The spectator passage 
time (estimated in sharp cut-off geometry) first increases and then remains
essentially constant as $b$ increases over the range of interest. On the other
hand, the expansion time decreases with increasing $b$ due to a decrease in~$d$. 
It is this decrease in the expansion time coupled with an essentially constant
passage time, which provides the driving force for more matter to escape 
the interaction region as $b$ is increased i.e. an increase in ``squeeze-out" with $b$.
The magnitude of the ``squeeze-out" follows an approximately linear
dependence because $d$ is roughly proportional to $1/b$ for the Au~+~Au impact parameter 
range 1~-~8 fm.

        At 6 AGeV the spectator passage time is very short
compared to the expansion time and preferential in-plane emission dominates.
In this case, however, the linear increase of $v_2$ with
increasing impact parameter is strongly influenced by the initial 
spatial asymmetry of the nuclei overlap region or participant matter. This
asymmetry is commonly characterized in terms of the width $L_x$ and
height $L_y$ of the overlapping region via
$\alpha_s=(L_y-L_x)/(L_y+L_x)$\cite{oll92} and can be shown to
be nearly linearly proportional to the impact parameter for medium
b values. The essentially flat dependence of $v_2$ observed at
4 AGeV suggests that, at the transition energy, the reduction in
the expansion time [in competition with the spectator passage time] with 
increasing~$b$, is compensated for by the (later) increased 
in-plane-emission from the preserved initial spatial asymmetry.

        The solid circles, open squares and solid triangles shown in
Figs.~\ref{bdep-v2}, represent results from calculations with a recent  
version of the Boltzmann Equation Model BEM\cite{dan98} which
assumes a soft ($K=210$~MeV), a stiff ($K=380$~MeV) and an intermediate
($K=300$~MeV) EOS respectively. The calculations include momentum dependent 
forces\cite{dan00}. A comparison of the
calculated $v_2$ values indicate sizeable differences between the
predictions for a stiff and a soft EOS for all three beam
energies. For both 2 and 4 AGeV this distinction increases with increasing 
impact parameter indicating that the impact parameter dependence of elliptic 
flow lends a new and important constraint for the EOS.
At 2 AGeV, the $v_2$ values for the stiff EOS show good 
agreement, both in magnitude and trend, with the experimental data. 
At 4 AGeV the measured $v_2$ values lie between the calculated
result for a stiff and a soft EOS, and appear to be in better overall agreement with 
an intermediate form of the EOS. At 6 AGeV the data 
is less compatible with a stiff EOS, but does not allow a clear distinction 
between the soft and the intermediate ($K=300$~MeV) EOS.
The results of these comparisons can be taken as
being suggestive of a softening of the EOS as previously
reported in Ref.\cite{pin99}. However, it is
interesting to investigate whether or not the differential flow 
measurement $v_2(b, p_T)$ provides further constraints.
This line is pursued below.

\begin{figure}
\centerline{\epsfysize=5.2in \epsffile{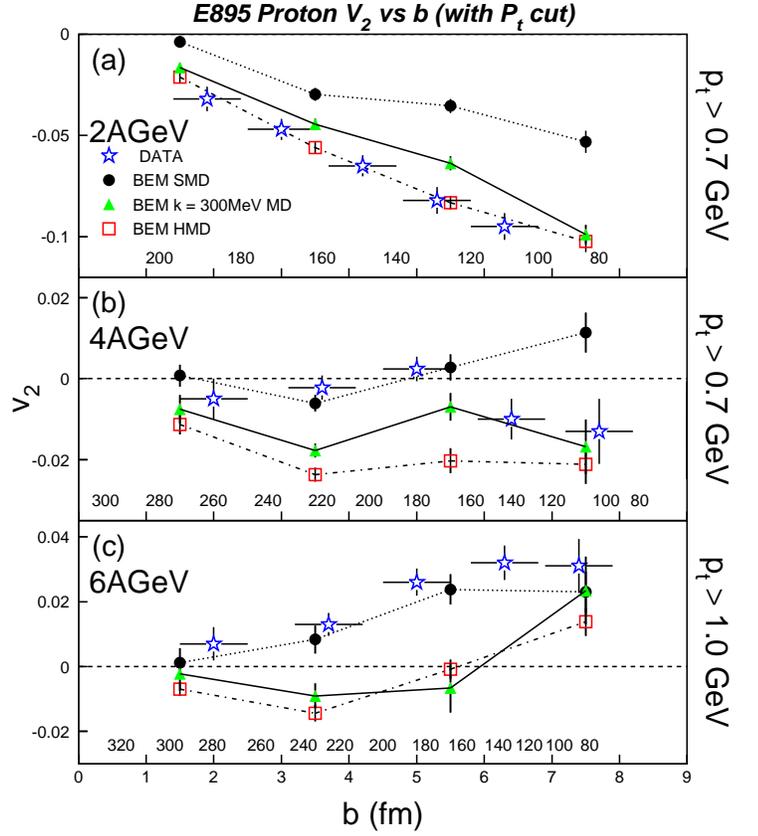}}
\vspace*{-0.2in}
\caption{Same as Figs.\ref{bdep-v2} except that a $p_T > 0.7 GeV$ cut
has been applied on the data and calculations at 2 and 4 AGeV and a $p_T > 1.0
 GeV$ cut has been applied at 6 AGeV.
}
\label{bdep-v2-ptcut}
\end{figure}

Figs.~\ref{bdep-v2-ptcut} compares experimental (stars) and calculated 
(circles, triangles and squares) differential elliptic flow $v_2(b, p_T)$ for 2, 4 
and 6 AGeV as indicated. At each beam energy, the BEM
calculations have been carried out for the same $p_T$ and $b$ selections 
applied to the data. Fig.~\ref{bdep-v2-ptcut} indicates good agreement between the 
data and the calculated results for a stiff EOS at 2 AGeV. At 4 AGeV the data again 
shows better overall agreement with the intermediate and soft EOS. At 6 AGeV the comparison 
indicates quite good agreement (both in magnitude and
trend) between the data and the results from the calculations
which assume a soft EOS. The latter agreement is in contrast to the results obtained from 
the comparison made in Fig.~\ref{bdep-v2}, and clearly indicates that the 
differential flow $v_2(b, p_T)$, does indeed provide an additional constraint for 
making a relatively clear distinction between the different EOS's at 6 AGeV. 


To summarize, we have studied differential proton elliptic flow 
in 2~-~6 AGeV Au~+~Au collisions.
The elliptic flow shows an essentially
linear dependence on $b$, in the range $1.5 \lesssim b \lesssim 8$ fm, with a
negative slope at 2 AGeV, approximately zero slope at 4 AGeV, and a positive
slope at 6 AGeV. This dependence can be understood in terms of (a) the 
relationship between the collision geometry,
(b) the relative magnitude of the time for
development of the transverse expansion, and (c) the
passage time for removal of the shadowing of participant
hadrons by the projectile and target spectators.
Detailed comparisons between the measured differential elliptic flow $v_2(b, p_T)$, 
and $v_2(b)$, and the results obtained from a relativistic Boltzmann-equation calculation 
not only suggest a softening of the EOS but also indicate that differential 
flow measurements provide very important constraints for the determination of 
the EOS of high density nuclear matter.

\acknowledgements

This work was supported in part by the U.S.~Department of
Energy under grants DE-FG02-87ER40331.A008, DE-FG02-89ER40531,
DE-FG02-88ER40408, DE-FG02-87ER40324, and contract DE-AC03-76SF00098; by the
US~National Science Foundation under Grants
No.~PHY-98-04672, PHY-9722653,
PHY-0070818,
PHY-9601271, and PHY-9225096; and by the University of Auckland Research
Committee, NZ/USA Cooperative Science Programme CSP 95/33.

%


\begin{tabular}{|l|ccc|}  \hline
            &             & Dispersion Correction Factor &   \\ \hline
b range (fm)& 2 AGeV      & 4 AGeV          & 6 AGeV  \\ \hline\hline
$0 < b < 3$ & 1.71         & 2.64             & 4.65     \\  \hline
$4 < b < 6$ & 1.22         & 1.59             & 2.47     \\  \hline
$7 < b < 8$ & 1.26         & 1.99             & 2.86    \\  \hline \hline
\end{tabular}

\vskip .3cm

{ Table 1:  Correction factors for reaction plane dispersion for several 
impact parameter ranges for the 2, 4 and 6 AGeV beam energies.
}

\end{multicols}

\begin{references}

\bibitem{stocker86} H. Stocker and W. Greiner, Phys. Rep. { 137},
                277, (1986).

\bibitem{qm96}
{\it Quark Matter '96, Proc.\ 12th Int.\
Conf.\ on Ultra-Relativistic Nucleus-Nucleus Collisions,
Heidelberg, Germany, 1996,} ed.\ P.\ Braun-Munzinger {\em et al.},
Nucl. Phys. { A610}, 1c--572c (1996).

\bibitem{reisdorf97} W. Reisdorf and H.~G.~Ritter, Annu. Rev.
Nucl. Part. Sci.  47, 663 (1997).

\bibitem{sor97}
H.\ Sorge, Phys.\ Rev.\ Lett.\ 78, 2309 (1997).

\bibitem{bar97}
J.~Barrette {\em et al.}, \prc{56}, 3254 (1997).

\bibitem{dan98}
P.~Danielewicz~{\em et al.},
\prl{81}, 2438 (1998).


\bibitem{pin99}
     C.\ Pinkenburg {\em et al.}, \prl{83}, 1295 (1999).

\bibitem{dan95}
P.~Danielewicz, Phys.\ Rev.\ C {51}, 716 (1995).

\bibitem{oll92}
J.-Y.\ Ollitrault, Phys.\ Rev.\ D 46, 229 (1992).
%

\bibitem{beam_energy}
Actual beam energies are 1.85, 3.9 and 5.9 AGeV respectively.

\bibitem{oll93}
J.-Y.\ Ollitrault, Phys.\ Rev.\ D 48, 1132 (1993).
%
\bibitem{GRai90} G. Rai, IEEE Trans. Nucl. Sci. 37,
56 (1990).
%
\bibitem{Bauer97} G. Bauer  , NIM A  386,  249
(1997).

\bibitem{dem90}
M.~Demoulins {\em et al.}, Phys.\ Lett.\ B 241, 476 (1990);
M.~Demoulins, Ph.\ D.\ Thesis, University Paris
Sud, 1989 (Report CEA-N-2628, CEN Saclay, 1990).

\bibitem{dan85}
P.~Danielewicz and G.~Odyniec, Phys.\ Lett.\ { B 157},
146 (1985).

\bibitem{gut89}
H.\ H.\ Gutbrod  , Rep.\ Prog.\ Phys.\ 52, 1267 (1989).

\bibitem{pid_mixup} Unique separation of $\pi^+$ and
protons was not achieved for all rigidities at all beam
energies.

\bibitem{oll97}
J.-Y.\ Ollitrault, nucl-ex/9711003 v2.

\bibitem{dan00}
P.\ Danielewicz, Nucl.Phys.A {673}, 375 (2000).

\end{references}
\end{document}